\documentclass[aps,pra,twocolumn,showpacs,floatfix]{revtex4}
\usepackage{bm}
\usepackage{graphicx} 
\usepackage{amsmath}
\usepackage{braket}
\begin{document}
\title{Non-Sticking of Helium Buffer Gas to Hydrocarbons}
\author{James F. E. Croft}
\author{John L. Bohn}
\affiliation{JILA, NIST, and Department of Physics, University of Colorado, Boulder, Colorado 80309-0440, USA}

\date{\today}

\begin{abstract}
Lifetimes of complexes formed during helium-hydrocarbon collisions at low temperature are estimated for symmetric top hydrocarbons. The lifetimes are obtained using a density-of-states approach. In general the lifetimes are less than 10-100~ns, and are found to decrease with increasing hydrocarbon size. This suggests that clustering will not limit precision spectroscopy in helium buffer gas experiments. Lifetimes are computed for noble-gas benzene collisions and are found to be in reasonable agreement with lifetimes obtained from classical trajectories as reported by Cui {\it et al} \cite{Cui:2014}.
\end{abstract}


\maketitle
\section{Introduction}

In the gas phase, no two molecules can truly stick upon colliding, unless there exists some mechanism for releasing their binding energy.  Most often, this mechanism involves a third molecule, hence molecule clustering proceeds more slowly as the density of the gas is reduced.  Thus in molecular beam expansions, the occurrence of seed gas atoms (typically noble gasses) adhering to entrained molecular species can be controlled by varying the pressure of the expansion.  Likewise, reduced three-body recombination rates in extremely rarefied ultracold gases are what allow phenomena such as Bose-Einstein condensation to be studied at all.

More precisely, three-body recombination, scaling as $n^3$ (where $n$ is the number density), dwindles in comparison to the two-body collision rate which scales as $n^2$.  A complication occurs when the two-body collisions result in the formation of a collision complex with sufficiently long lifetime $\tau$ that collision with another molecule can occur during the interval $\tau$,  resulting in a bound state of two particles and a release of energy, a process  known as the Lindemann mechanism \cite{Levine:molecular:2005}.
The importance of this mechanism relies crucially on the scale of $\tau$ for a given low-density gas.  Recently it has been suggested -- though not yet empirically verified -- that $\tau$ can grow quite long for extremely low-temperature gases, even of order 100 ms for alkali-metal dimer molecules colliding in microKelvin gases \cite{Mayle:2013}.  If true, this process would lead to the decay of such a gas.  

Between the regimes of supersonic jet expansions and ultracold molecules lies a host of experiments on cold molecules, notably, those cooled to the temperature of an ambient helium buffer gas in a cold cell at temperatures on the order of 1-10~K.  The buffer-gas cell has proven to be a reliable source of cold molecules, at temperatures sufficiently low to extend the reach of precision spectroscopy \cite{Patterson:enantiomer:2013,Naima:LiHe:2013,Shubert:Chiral:2014,ACME:2014}.  The success of these experiments requires the helium buffer gas not to stick to the molecules under study, as the spectral lines would thereby be shifted.  Under typical experimental conditions, the collision rate of the buffer gas with the molecule is tens of $\mu$sec \cite{Piskorski:2014}; spectroscopy is safe if the collision complex lives for shorter times than this.  

Thus far, no empirical evidence has emerged suggesting that the sticking occurs in the buffer gas environment \cite{Patterson:2010,Patterson:2012,Patterson:enantiomer:2013,Piskorski:2014,Piskorski:Thesis:2014}, a conclusion that is supported by detailed classical trajectory calculations \cite{Li:2012,Li:2014,Cui:2014}. This appears to be true even for a relatively floppy molecule such as trans-stilbene, where comparatively low energy vibrational modes might have been expected to promote sticking \cite{Piskorski:2014}. The existing evidence suggests, therefore, that  the transient lifetime of a hydrocarbon-helium complex in the buffer gas cell remains comfortably less than $\sim \mu$sec. 

In this article we argue that such short lifetimes are natural and perhaps even generic under these circumstances.  The argument is based on considerations drawn from the theory of unimolecular dissociation, in which a complex molecule with sufficient energy to dissociate nevertheless experiences a time delay before actually doing so. In this theory, the dwell time of the complex stands at the balance between excitation of degrees of freedom that cannot dissociate while conserving energy (thus contributing to longer dwell times), and degrees of freedom that can (thus contributing to shorter dwell times).  For complexes consisting of a hydrocarbon molecule with a transiently attached helium atom, both these densities of states may increase with increasing molecule size, so that the dwell time $\tau$ depends weakly on the specific hydrocarbon. Based on simple ideas, we give order-of-magnitude estimates for typical lifetimes in such a gas.

\section{Buffer gas Environment}

Here we contemplate a buffer gas cell, in thermal equilibrium at temperature $T$, containing helium gas with number density $n_a$ and hydrocarbon molecules with density $n_m \ll n_a$, in which case the majority of collisions the molecules suffer will be with atoms.  Upon introducing molecules into the gas, atom-molecule collisions occur at a rate $K_{am} n_a n_m$, defined by a rate constant $K_{am}$.  In the Lindemann model , these collisions produce short-lived complexes that are characterized by number density $n_c$, and that decay at a mean rate $\gamma = 1/\tau$.  Under these circumstances the atomic density is not significantly depleted, and the collisions are described by the rate equations
\begin{eqnarray}
{\dot n}_m &=& -K_{am} n_a n_m + \gamma n_c \nonumber \\
{\dot n}_c &=& K_{am} n_a n_m - \gamma n_c.
\end{eqnarray}
After an equilibration time $\sim (K_{am}n_a)^{-1}$, the fraction of molecules temporarily absorbed in complexes is
\begin{eqnarray}
\frac{ n_c^{eq} }{ n_m^{eq} } \approx K_{am}n_a \tau,
\end{eqnarray}
a fraction that is negligible unless the dwell time $\tau$  is at least comparable to the inverse collision rate.

To place approximate numbers to this constraint, consider a typical helium number density of $n_a = 2 \times 10^{14}$ cm$^{-3}$ \cite{Piskorski:2014}, and a collision cross section approximated by the Langevin capture cross section \cite{Levine:molecular:2005},
\begin{eqnarray}
\sigma_L = \pi \left( \frac{ 3}{ 2 } \right)^{2/3} \left( \frac{ 6 C_6 }{ 2 k_BT } \right)^{1/3} \approx 3 \times 10^{-14}  {\rm cm}^2,
\end{eqnarray}
 assuming a van der Walls coefficient of $C_6 = 100$ atomic units (see below).  The atom-molecule rate constant is then $K_{am}{\bar v} \sigma_L \approx 6 \times 10^{-10}$ cm$^3$/s. whereby the fraction of complexes is approximately
 \begin{eqnarray}
\frac{ n_c^{eq} }{ n_m^{eq} } \approx \frac{ \tau }{ 10 \mu{\rm s} }.
\end{eqnarray}
Thus for dwell times significantly less than 10$\mu$s, the complexes should be rare.  In what follows, we estimate the lifetime, finding it to be at most 10-100 ns.  Therefore, in the buffer gas we expect fewer (probably far fewer) than one molecule in a hundred to be involved in a collision complex at any given time.  

\section{Rates and Lifetimes}
We are interested here in identifying an upper bound for the sticking lifetime of helium atoms on hydrocarbon molecules.  The sticking process is denoted schematically as
\begin{eqnarray}
\label{eqn:RateEqn}
\rm{He} + \rm{M}(X) \rightarrow (\rm{He}+\rm{M})^*(JM_J) \xrightarrow{\tau(J,M_J)} \rm{He} + \rm{M}(X')
\end{eqnarray}
where $X$ are a set of quantum numbers, including molecular rotation $N$, which completely describe the state of the molecule.  For the duration of the collision, the atom and molecule are assumed to reside in a complex with total angular momentum $J$.  This angular momentum is regarded as the vector sum, in the quantum mechanical sense, of the molecule's rotation $N$ and the partial wave of the atom-molecule relative motion, $L$.  
 $\tau(J,M_J)$ is the lifetime of a complex for total angular momentum $J$ and projection $M_J$.  

At a given collision energy $E_c$, collisions can occur in any of a set of incident channels, whose number is the number of energetically open channels, $N_0(J,M_J)$, for a given total angular momentum. 
The relevant mean sticking lifetime in the experiment is therefore the lifetime of each collision complex averaged over all $J$ and $M_J$ combinations, and weighted by the number of incident channels leading to that combination:
\begin{eqnarray}
\label{eqn:average_tau}
\bar{\tau} = \frac{\sum_{J,M_J}\tau(J,M_J)N_o(J,M_J)}{\sum_{J,M_J}N_o(J,M_J)}
\end{eqnarray}
Within the  Rice-Ramsperger-Kassel-Marcus (RRKM) model \cite{Marcus:Vol1:1952,Marcus:Vol2:1952,Levine:molecular:2005} the dwell time of a complex is approximated as
\begin{eqnarray}
\label{eqn:rrkm}
\tau(J,M_J) = 2 \pi  \hbar \frac{  \rho(J,M_J)}{N_o(J,M_J)},
\end{eqnarray}
where $\rho(J,M_J)$ is the density of available ro-vibrational states (DOS) of the complex for the given total angular momentum. Thus the mean sticking lifetime is given by
\begin{eqnarray}
{\bar \tau} = 2 \pi \hbar \frac{ \sum_{J,M_J} \rho(J,M_J) }{ \sum_{J,M_J} N_o(J,M_J) }.
\end{eqnarray}
It must be emphasized that this approximation is an upper limit to the lifetime, as it assumes that all the possible states contributing to $\rho$ are in fact accessible to be populated in a collision.  This assumption disregards, for example, barriers in the potential energy surface that forbid a given entrance channel from probing a certain region of phase space.  This circumstance would reduce the effective density of states, hence also the lifetime.

\section{Density of states}
Following Mayle {\it et al} \cite{Mayle:2012,Mayle:2013}  we estimate the density of states $\rho(J,M_J)$ by a counting procedure. This begins by somewhat artificially separating the degrees of freedom of the He-molecule complex into those coordinates $\{ {\bf X} \}$ necessary to describe  internal motions of the molecule, and a coordinate ${\bf R}$ giving the relative motion of the atom and molecule.  The enumeration of molecular states follows from the spectrum of the molecule.  The atom-molecule states are approximated by postulating a schematic atom-molecule potential $V(R)$.  

For a given molecular state with energy $E(X)$, the potential $V_{X,L}(R) = V(R) + \hbar^2 L(L+1)/2\mu R^2 + E(X)$ is constructed, so far as $L$ and the molecular rotation are consistent with the total angular momentum $J$ under consideration.  The number of bound states $N_{am}(X,L)$ of $V_{X,L}$, lying within an energy range $\Delta E$, centered around the collision energy, is found.  The density of these states  is then given by the sum
\begin{eqnarray}
\rho(J,M_J) = \frac{ 1 }{ \Delta E } \sum_{X,L}^{\prime}N_{am}(X,L).
\end{eqnarray}
The prime on the summation sign is a reminder that the sum is taken over those quantum numbers for which energy and angular momentum conservation are satisfied.

\begin {table}[bp]
\begin{center}
\begin{tabular}{| l || c | c | c | c |} 
  \hline 
  System &  $R_e$~(\AA) & $V^{min}$~(K) & $C_6$~(au) & $\rho_{am}$~(K$^{-1}$) \\
  \hline  
  \hline                       
Helium+Methane     &  3.2 & 52  & 16 & 0.05\\
Helium+Ethane      &  3.3 & 77  & 27 & 0.03\\
Helium+Propane     &  3.7 & 82  & 62 & 0.03\\
Helium+Butane      &  3.5 & 108 & 59 & 0.02\\
Helium+Pentane     &  3.2 & 126 & 41 & 0.02\\
Helium+Hexane      &  3.4 & 131 & 55 & 0.02\\
\hline
Helium+Benzene     &  3.0 & 130 & 30 & 0.02\\
Helium+Naphthalene &  3.2 & 159 & 45 & 0.01\\
\hline
Helium+Propandiol  &  3.4 & 115 & 56 & 0.02\\
\hline
\end{tabular}
\caption{\label{tab:OptimizedData} Equilibrium distance and energy minimum for the helium-hydrocarbon interaction for a variety of different systems. Equilibrium distance and potential were obtained in GROMACS \cite{Hess:GROMACS:2008} with the OPLS-AA force field \cite{Jorgensen:OPLSAA:1996,Kaminski:OPLSAA:2001}.  In each case, these data can be used to construct a schematic Lenard-Jones potential $V(R)$, leading to the atom-molecule density of states factor $\rho_{am}$, defined in (\ref{eqn:rho_am}). }
\end{center}
\end {table}

In the model, the potential $V(R)$ is assumed to be of Lennard-Jones form.  This potential has a realistic van der Waals coefficient for the He-hydrocarbon interaction, as well as a reasonable minimum.  A key point in the lifetime analysis is that the parameters of this potential are  {\it weakly dependent} on the particular hydrocarbon involved. Table \ref{tab:OptimizedData} shows the equilibrium distance, van der Waals coefficient, and energy minimum for a variety of Helium-Hydrocarbon systems. While this table comprises a variety of hydrocarbons of different shapes and sizes, the equilibrium distance and energy minimum vary little between them. This is because the helium atom only interacts with the nearby atoms in the hydrocarbon. Further, the reduced mass for the collision of a hydrocarbon with helium is to a very good approximation simply the helium mass. 

Thus the potentials $V_{X,L}(R)$, and the numbers of states $N_{am}(X,L)$ that they hold, vary little between different helium-hydrocarbon systems. We therefore make the approximation
\begin{eqnarray}
\label{eqn:rho}
\rho(J,M_J) \le \rho_{am} \sum_{X,L}^{\prime} 1  \equiv \rho_{am}N_{m}(J,M_J),
\end{eqnarray}
where $N_m(J,M_J)$ is the number of states of the molecule consistent with angular momentum and energy conservation.  The quantity $\rho_{am}$ is a kind of representative atom-molecule density of states.  It is conveniently approximated by the inverse of the lowest vibrational excitation,
\begin{eqnarray}
\label{eqn:rho_am}
\rho_{am} = \frac{1}{E_{v=1,L=0}-E_{v=0,L=0}}
\end{eqnarray}
Thus the  complex lifetime is approximately bounded above by
\begin{eqnarray}
\label{eqn:rrkm_approx}
{\bar \tau} \approx   2 \pi  \hbar \rho_{am} \times  \frac{ \sum_{J,M_J} N_{m}(J,M_J) }{ \sum_{J,M_J} N_o(J,M_J)}.
\end{eqnarray}
Here the first factor has a generic approximate value for any He-hydrocarbon interaction (see Table I), while the second factor elaborates on the distinction between different molecules.  Upon increasing the density of states of the molecule, both the numerator and the denominator of this factor could increase.  The counting exercise must be done to find its ultimate effect on the molecular lifetime.  

\section{Effect of rotational states}

\begin{figure}[tbp]
\centering
\includegraphics[width=0.99\columnwidth]{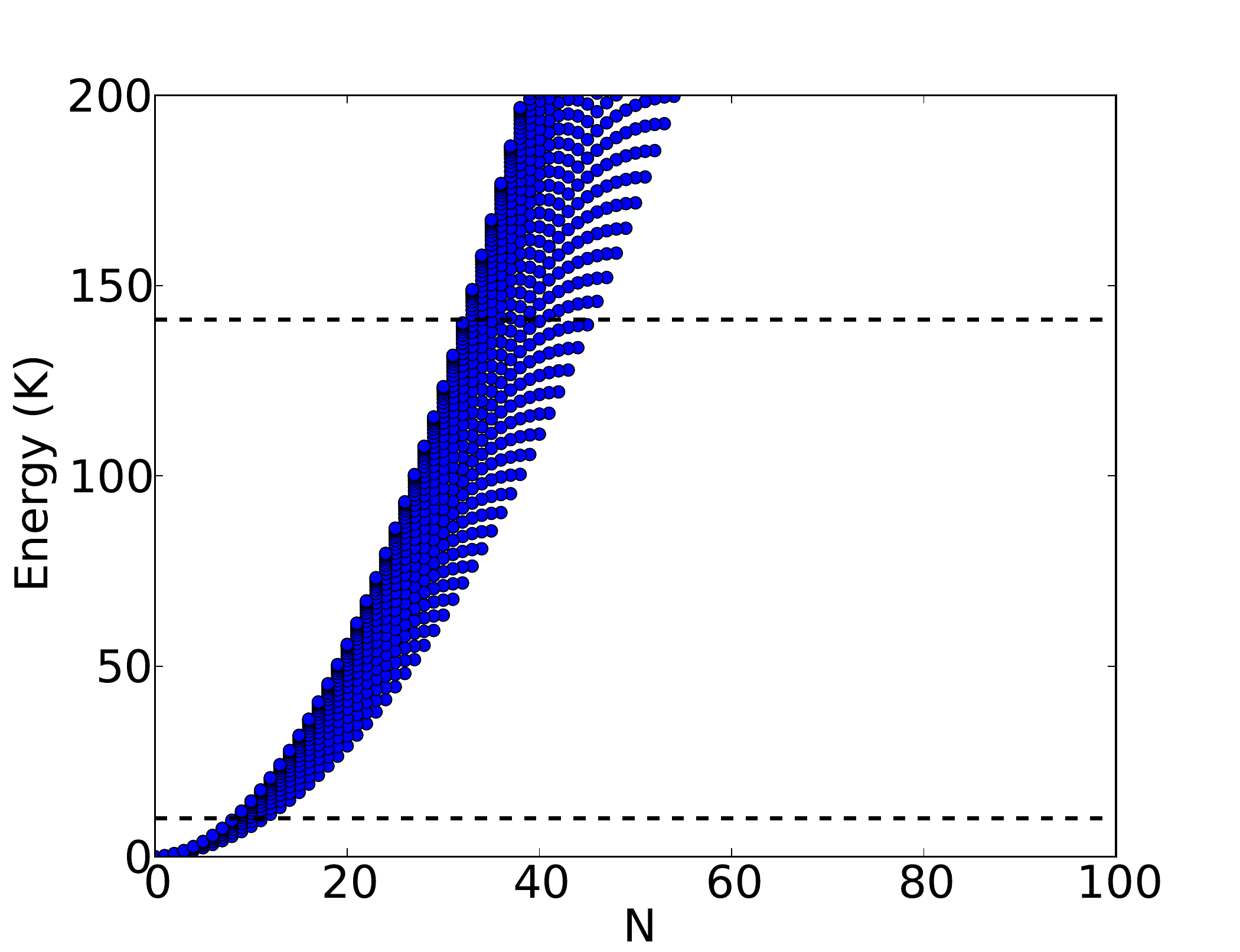}
\includegraphics[width=0.99\columnwidth]{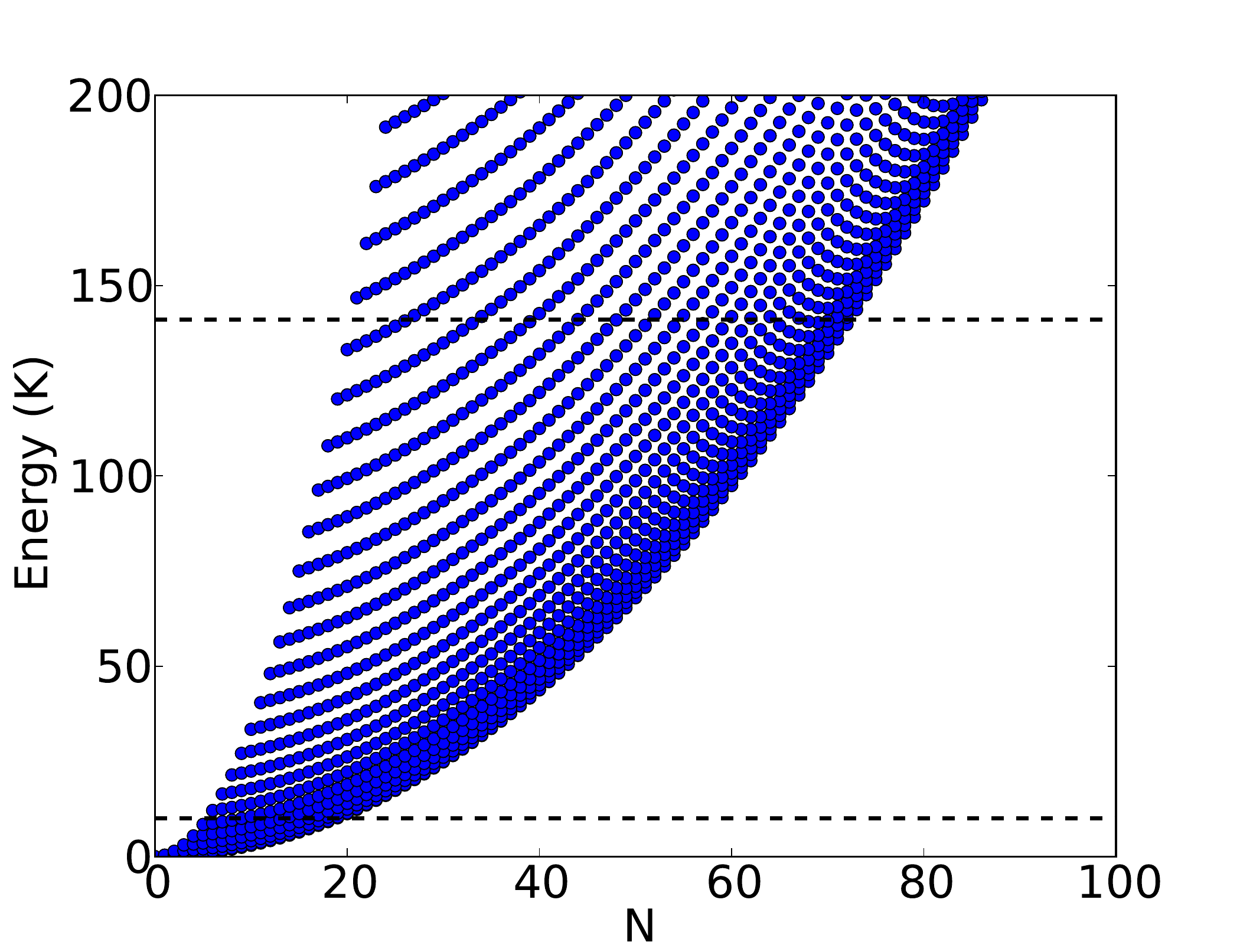}
\caption{(Color online) Rotational energy levels for hexane (upper panel) and benzene (lower panel). The dashed lines show the collision energy $E_c = 10$~K and the highest possible threshold energy that can contribute to the DOS, $|V_{L=0}(R_{min})|+E_c$.  While hexane has more levels above $E_c$ that can contribute to sticking, is also possesses more levels below $E_c$ that can lead to dissociation of the atom-molecules complex.} 
\label{fig:rotational_energy_levels}
\end{figure}

Rotational splittings in large hydrocarbons tend to be small compared to the collision energy in a buffer gas $\Delta E_{rot} \ll E_c \approx 10$~K  while in general the vibrational splitting is larger $\Delta E_{vib} > E_c$. The dominant contribution to $N_{m}$ and $N_o$, in equation (\ref{eqn:rrkm_approx}), arises therefore from  the rotational levels of the molecule. Figure \ref{fig:rotational_energy_levels} shows the lowest rotational energy levels for both hexane and benzene. The  rotational constants were obtained from the Computational Chemistry Comparison and Benchmark Database (CCCBD) \cite{CCCBD} and the energy levels computed with PGOPHER \cite{PGOPHER}. It is seen that these two systems have very different rotational energy levels. 

Shown in Figure \ref{fig:rotational_energy_levels} are two dashed lines.  The lower one is 10~K, the approximate collision energy at buffer gas temperatures.  The upper one represents the the collision energy plus the depth of the schematic potential $V(R)$ between the atom and the molecule, a quantity denoted $E_{max} = E_c + |V_{L=0}(R_{min})|$.  Ignoring for the moment considerations of angular momentum conservation, the total number of states belonging to any potential $V_{X,L}$ and lying in energy below $E_{max}$ denote potentially resonant states that contribute to increasing the lifetime; states higher in energy than this do not satisfy energy conservation.  

\begin{figure}[tbp]
\centering
\includegraphics[width=0.99\columnwidth]{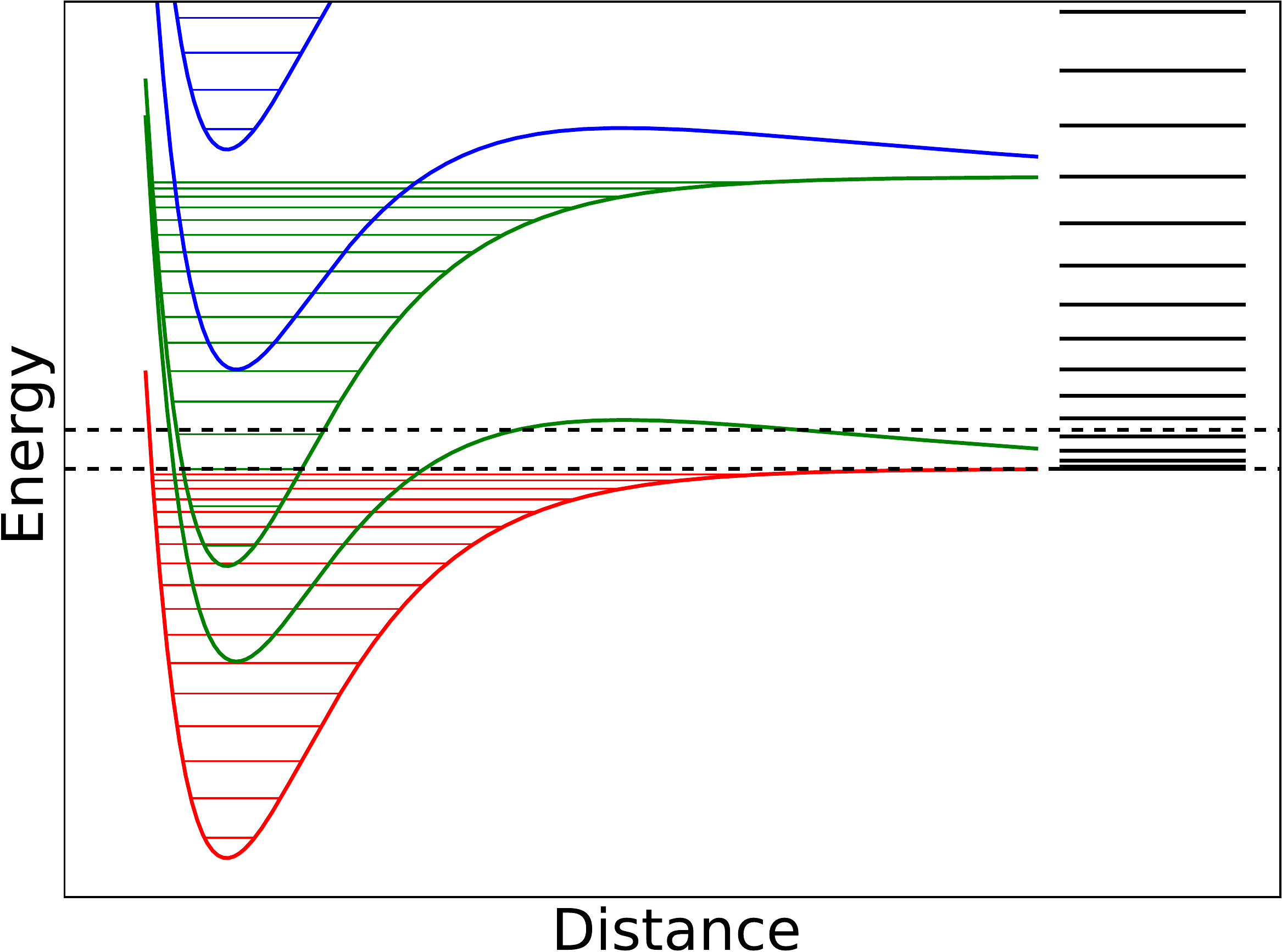}
\caption{(Color online) Schematic showing contributions to $N_m$ and $N_o$. The dashed horizontal lines represent zero energy and the collision energy. The horizontal solid black lines on the right represent rotational levels of the molecule $E(N)$. Combinations of $N$ and $L$ consistent with angular momentum conservation lead to potentials $V_{N,L}$, only a few of which are shown for clarity. Potentials $V_{N,L}$ whose minimum lies below the collision energy contribute to $N_m$; these potentials are colored green. Likewise potentials $V_{N,L}$ with a threshold and centrifugal barrier below the collision energy contribute to $N_o$ and are colored red. Potentials which cannot contribute to either are colored blue.} 
\label{fig:DOS_Schematic}
\end{figure}

In more detail, the number of states $N_m$ must be counted in a way consistent with the conservation of angular momentum.  Thus for a given fixed total $J$ the possible rotation $N$ and partial wave $L$ quantum numbers are considered, and the potentials $V_{N,L} = V(R) + \hbar^2 L(L+1)/2 \mu R^2 + E(N)$ constructed.  If the minimum of this potential lies below the collision energy, then this state is energetically allowed and is counted as part of $N_m$; otherwise not (see Figure \ref{fig:DOS_Schematic}). Likewise, if the centrifugal barrier of the potential $V_{N,L}$ lies below the collision energy, then the state is counted toward the number of open channels $N_o$.  Otherwise, the collision is assumed not to tunnel through this barrier and does not count as an entrance or exit channel.  This requirement is essentially the same as assumed in the Langevin capture model of collisions.

Computing the sum in this way, we find lifetimes of hexane and benzene to be approximately 36 ps and 44 ps, respectively.  Gratifyingly, the lifetime for benzene is consistent with the far more detailed classical trajectory calculations of Cui, Li, and Krems \cite{Cui:2014}.  

Within such a model, we can consider the lifetimes for many hypothetical molecules, characterizing their rotational spectra  by the symmetric top energy levels
\begin{eqnarray}
E(N,K) = BN(N+1) + (A-B)K^2.
\end{eqnarray}
These lifetimes assume, as above, that $\rho_{am}$ is approximately the same for all such molecules.  To make the calculation concrete, we assume the same value of $\rho_{am}$ and the schematic potentials $V_{L,N}$ as for hexane.

\begin{figure}[tbp]
\centering
\includegraphics[width=0.99\columnwidth]{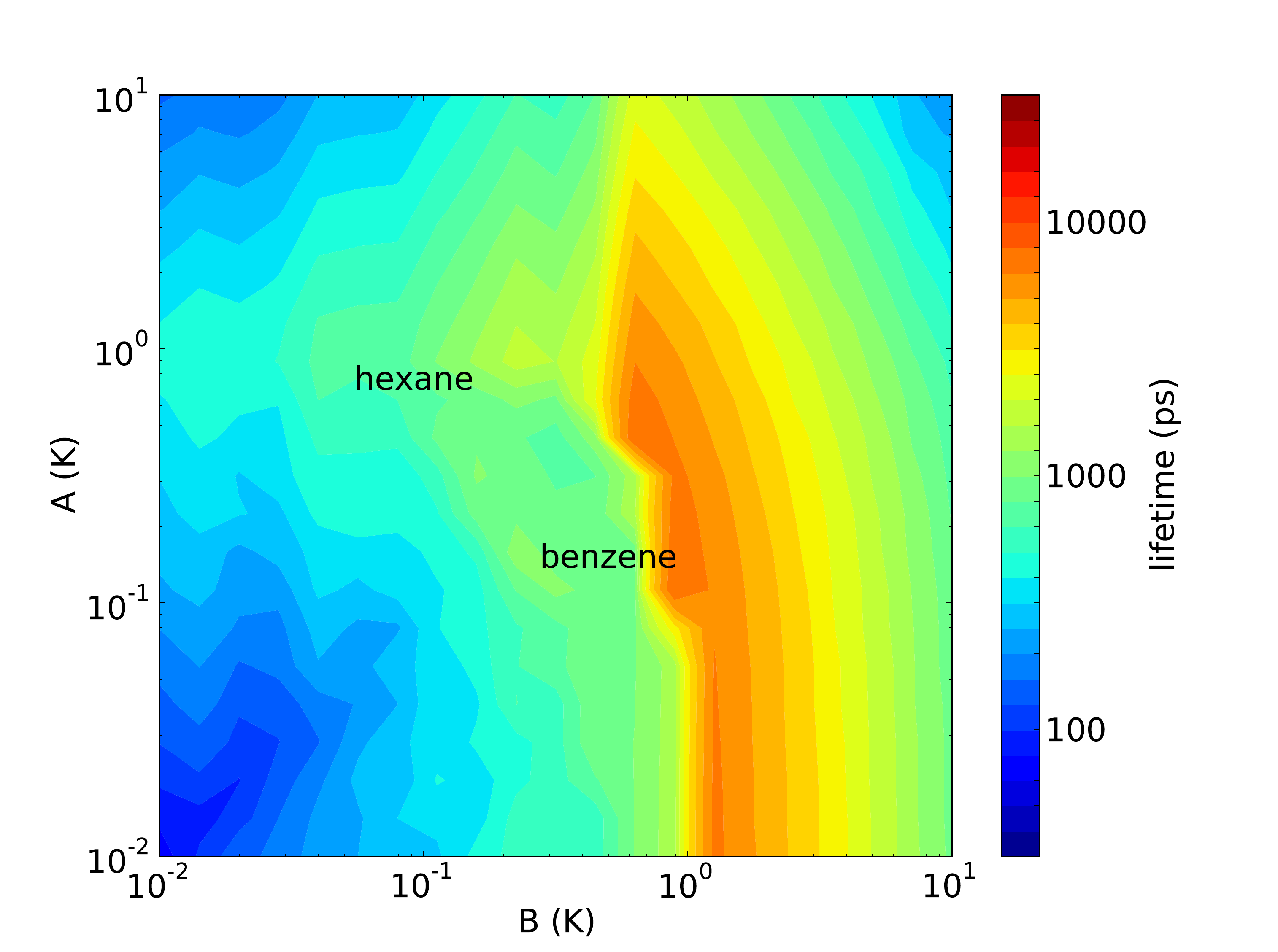}
\includegraphics[width=0.99\columnwidth]{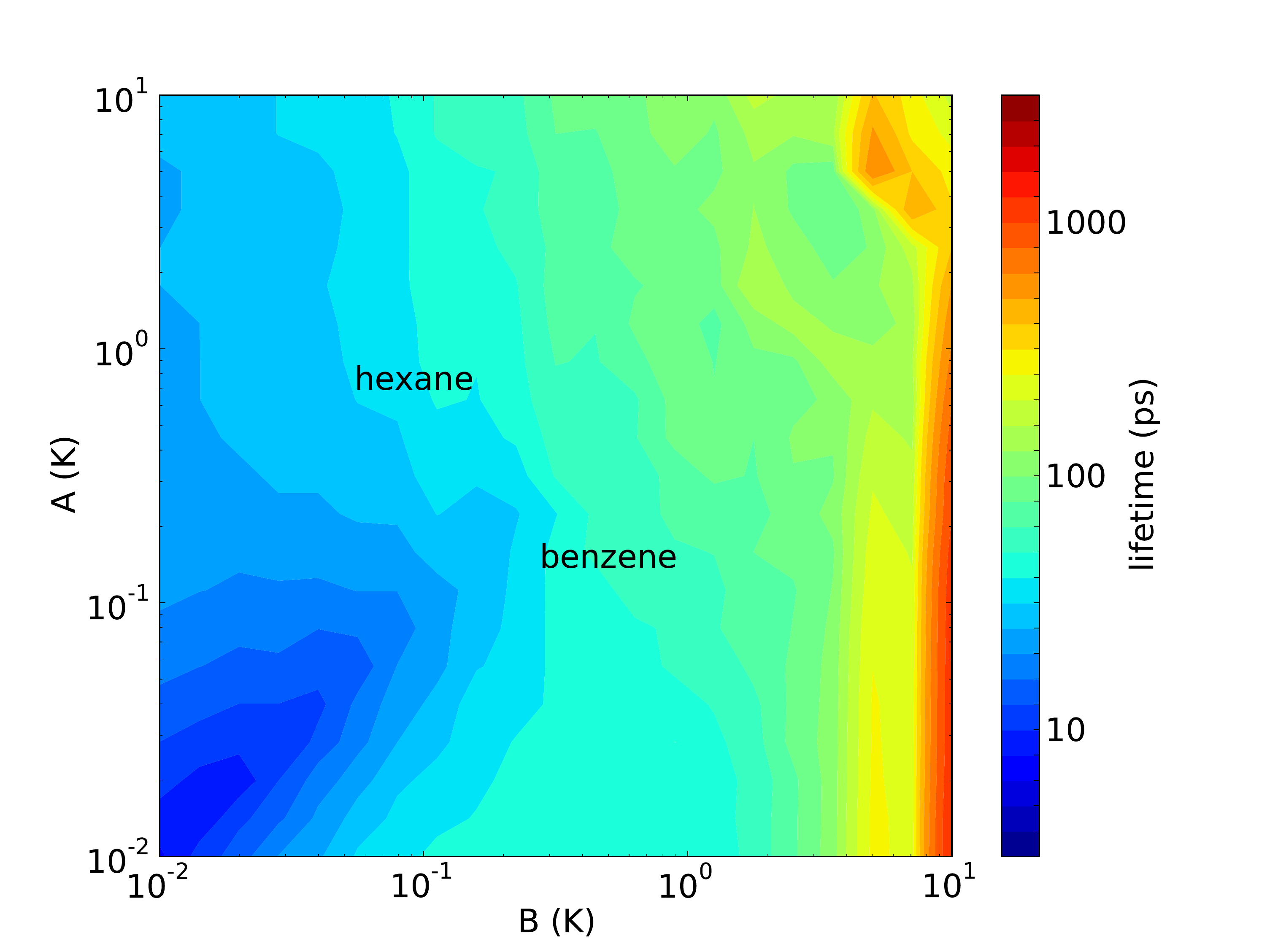}
\caption{(Color online) Lifetime of helium symmetric-top molecule clusters as a function of rotational constants $A$ and $B$ at $E_c=1$~K (upper panel) and $E_c=10$~K (lower panel). Using $\rho_{am}$ and $V_{L,N}$ as for hexane. Labeled are the lifetime for the prolate top hexane and oblate top benzene, the point labeled is at the bottom left of the word.} 
\label{fig:lifetime_contour}
\end{figure}

Figure \ref{fig:lifetime_contour} shows the lifetime of symmetric top molecules within this model, as a function of rotational constants $A$ and $B$, at two different collision energies,  $E_c=1$ and 10~K. The longest lifetimes, for both collision energies, occur when $B \approx E_c$. The lifetime is only weakly dependent on $A$. As $B$ decreases below $E_c$ the lifetime rapidly decreases, because in this circumstance rotational levels lying below $E_c$  contribute to $N_o$ in addition to $N_m$. In addition, as $B$ increases above $E_c$ the lifetime slowly decreases, as rotational levels are pushed higher and fewer contribute to $N_o$. We therefore conclude that maximum lifetimes occur when $B \gtrsim E_C$.  Finally, for a given molecular spectrum, the lifetimes are larger for lower collision energy $E_c$, since relatively more of the molecular states contribute to $N_m$ than to $N_o$. 

For buffer gas experiments at 10~K this means the maximum lifetime actually occurs for light species such as methane, where $B=7.6$~K. This lifetime is around 1~ns well below the 10~$\mu$s required for clustering to occur. This result is quite promising for the prospect of cooling large  hydrocarbons. It is also worth remembering that the RRKM lifetime is an upper-bound on the actual lifetime as it assumes ergodicity, an assumption that appears justified for helium-benzene collisions \cite{Cui:2014}.

These remarks are derived for symmetric top molecules, but little should change for asymmetric top molecules.  
The lifetime of a symmetric rotor at a given buffer gas temperature is primarily determined by its principal rotational constant $B$. The rotational energy levels of an asymmetric top are intermediate between prolate and oblate limits. As such it is expected that, as with symmetric tops, the lifetime of asymmetric tops will also be primarily determined by $B$.

Thus far we have considered lifetimes of collision complexes of helium with large hydrocarbons, but one may also contemplate buffer gas cooling large biological molecules such as nile-red \cite{Piskorski:2014}. In general such molecules have a hydrocarbon backbone with functional groups containing elements such as oxygen, nitrogen etc. While the interaction of helium with such elements can be stronger (compare propandiol and propane in table \ref{tab:OptimizedData}). this should be a minor effect.  Indeed, that the estimated lifetimes for propane and propandiol are 54 and 39~ps respectively. 

\section{Influence of vibrational states}
Including vibrational energy levels of the molecule will presumably increase the lifetime of the complex, by increasing the density of states $\rho$ without significantly increasing the number of open channels $N_o$ (this latter fact follows because the vibrational constant is likely to be larger than 10~K). The longest increase in  lifetime will occur when a vibrational level exists just above the collision energy $\Delta E_{vib} \gtrsim E_c \approx 10 $~K, so that it contributes to $N_m$ but not to $N_o$.  Even in this case,  perhaps ten vibrational levels would occur in the energy range up to $E_{max}$, meaning that the lifetimes could increase from the estimates in the previous section by perhaps an order of magnitude, up to tens to hundreds of nanoseconds at $E_c=10$K. This short lifetime is consistent with  the lack of clustering observed in trans-stillbene and nile red where low energy vibrational modes might have been expected to promote sticking \cite{Piskorski:2014}. 

\section{Alternative Noble gases}
\label{sec:NG}
Other noble gas (NG) atoms are potential candidates for buffer-gas cooling and supersonic-expansion experiments \cite{Patterson:2009}. Cui {\it et al} \cite{Cui:2014} have reported noble-gas benzene complex lifetimes, for temperatures in the range 5-10~K, from classical trajectory simulations. Table \ref{tab:NGBenzenelifetimes} compares the DOS lifetimes with those of Cui {\it et al} at 10~K. We compute lifetimes separately for each of the cross-sections of the NG-benzene potential reported in \cite{Cui:2014}. As for the classical trajectory lifetimes the DOS lifetime increases with NG mass, as deeper potentials lead to higher $N$ and $L$ quantum numbers contributing to $N_m$. 

The DOS lifetimes always overestimate the classical trajectory lifetime and by an amount increasing with the mass of the NG atom. The DOS lifetime assumes ergodicity, that is the full density of rotational states is actually populated in a collision. If this is not the case then the lifetime of the cluster is reduced. We interpret the increasing overestimation of the lifetime, with NG mass, as evidence that high rotational states available in the collision are not necessarily populated. Intuitively this can be understood as the the shape of the benzene hindering the rotation of the NG atom around it.  Nevertheless, the estimates for experimentally relevant helium buffer gas remain fairly accurate, in cases where the comparison can be made.  
\begin {table}[bp]
\begin{center}
\begin{tabular}{| l || c | c | c | c |}
  \hline  
  & \multicolumn{4}{|c|}{$\tau$~(ps)} \\
  \hline 
  System &  Out-of-plane & Vertex-in-plane & Side-in-plane & Cui {\it et al} \\
  \hline  
  \hline                       
Helium    &  40   & 40   &  30   & $\sim$10   \\
Neon      &  260  & 240  &  140  & $\sim$50   \\
Argon     &  1380 & 1080 &  640  & $\sim$100  \\
Krypton   &  2900 & 1990 &  1230 & $\sim$150  \\
Xenon     &  4470 & 2990 &  1920 & $\sim$200  \\
\hline
\end{tabular}
\caption{\label{tab:NGBenzenelifetimes} Lifetimes in picoseconds for noble-gas benzene complexes. For both DOS lifetimes using cross-sections of the NG-benzene potential and classical trajectory lifetimes reported in \cite{Cui:2014}.}
\end{center}
\end {table}

\section{Conclusions}
In the present work we have developed a method for estimating helium-hydrocarbon complex lifetimes, using a density-of-states approach, at low collision energies. This model distinguishes between degrees of freedom that do not have energy to dissociate (contributing to longer lifetimes), and degrees of freedom that do (contributing to shorter lifetimes). The lifetime of a complex is determined by the balance between these. We obtain lifetimes for generic symmetric-top hydrocarbons finding that the lifetime decreases with increasing hydrocarbon size. This result is extremely encouraging for using helium as a buffer gas for cooling large biological molecules, which relies on helium buffer gas not to stick to the molecules. This result is in agreement with all empirical evidence \cite{Patterson:2010,Patterson:2012,Patterson:enantiomer:2013,Piskorski:2014,Piskorski:Thesis:2014} and other theoretical calculations \cite{Li:2012,Li:2014,Cui:2014} based on classical trajectories. Our approach complements these, enabling a rough survey of molecular species and their behavior in the buffer gas environment.

Finally, we note that in some case the lifetimes are not always many orders of magnitude below 10$\mu$sec, but in some cases may be as high as tens to hundreds of nanoseconds.  Moreover, lifetimes increase at lower collision energies, while collision rates increase at higher buffer-gas densities.  Thus sticking may be an observable effect, in slightly colder, denser helium cells, for well-chosen molecules.

\begin{acknowledgments}
This work was supported by the Air Force Office of Scientific Research under the Multidisciplinary University Research Initiative Grant No. FA9550-1-0588.  We acknowledge useful conversations with J. Piskorski, D. Patterson, and J. Doyle.
\end{acknowledgments}

\bibliography{jmh_all,jfec_all}

\end{document}